\documentclass[twocolumn,aps,prl,preprintnumbers,showpacs,nofootinbib]{revtex4}
\usepackage[utf8]{inputenc}
\usepackage[english]{babel}
\usepackage{amsmath}
\usepackage{amsfonts}
\usepackage{amssymb}
\usepackage{epsfig}
\usepackage{graphics,psfrag,rotating}
\usepackage{graphicx}
\usepackage{dcolumn}
\usepackage{bm}
\usepackage{epstopdf}
\usepackage{color}
\usepackage[usenames,dvipsnames,svgnames]{xcolor}
\usepackage[colorlinks=true,
            linkcolor=red,
            urlcolor=gray,
            citecolor=blue]{hyperref}
  \usepackage{hyperref}

\newcommand{\lsim}   {\mathrel{\mathop{\kern 0pt \rlap
{\raise.2ex\hbox{$<$}}}
 \lower.9ex\hbox{\kern-.190em $\sim$}}}
\newcommand{\gsim}   {\mathrel{\mathop{\kern 0pt \rlap
{\raise.2ex\hbox{$>$}}}
\lower.9ex\hbox{\kern-.190em $\sim$}}}

\def\3nab{\tilde{\nabla}}

\def\hsp5{\hspace{5mm}}

\def\case#1/#2{\textstyle\frac{#1}{#2}}

\def\ber {\begin{eqnarray}}
\def\eer {\end{eqnarray}}
\def\bea {\begin{eqnarray}}
\def\eea {\end{eqnarray}}

\def\bc {\begin{center}}
\def\ec {\end{center}}
\def\case#1/#2{\frac{#1}{#2}}

\newcommand{\bw}{\begin{widetext}}
\newcommand{\ew}{\end{widetext}}

\newcommand{\be}{\begin{equation}}
\newcommand{\bse}{\begin{subequation}}
\newcommand{\ese}{\end{subequation}}
\newcommand{\ee}{\end{equation}}
\newcommand{\eei}{\end{eqnarray}\indent\indent}
\newcommand{\ba}{\begin{array}}
\newcommand{\ea}{\end{array}}
\newcommand{\bal}{\begin{eqnarray}}
\newcommand{\eal}{\end{eqnarray}}

\def\case#1/#2{\textstyle\frac{#1}{#2} }

\newcommand{\bver}{\begin{verbatim}}
\newcommand{\ever}{\end{verbatim}}

\begin{document}


\title{ Cosmography of chameleon gravity}
\author{
A. Salehi\footnote{Email: salehi.a@lu.ac.ir},
}
\affiliation{ Department of Physics, Lorestan University, Khoramabad, Iran}

\date{\today}

\begin{abstract}
The paper discusses a particular model of chameleon gravity where the scalar field has an exponential as $V=M^{4}\exp(\frac{\alpha\phi}{M_{pl}})$ and an exponential coupling to the matter density component of the Universe as $\rho_{m}\exp(\frac{\beta\phi}{M_{pl}})$. We finds the relations between the four (in principle free) parameters of the model and the first five cosmographic parameters (which can be written in terms of the Hubble parameter and its first four derivatives). If such cosmographic parameter are known then the we present the algebraic relations which can be used to derive the free parameters in terms of the cosmographic ones. We show that, determining the third derivative of scale factor $Q$ is of great importance if the chameleon parameter $\beta$ and coupling constant $\alpha$ are known. On the other hand, if nothing can be proposed about $(\alpha,\beta)$, reconstructing the free parameters of the model
 can be done by measuring the first five derivatives of the scale factor and testing the validity of theory is only possible by its first six derivatives, so that the parameters $(\alpha,\beta)$ will also be determined automatically. We also find relation between cosmographic parameters and parameters of the model when it dynamically approaches its critical points.
\end{abstract}


%
%


\maketitle
In 1970, Alan Sandage \cite{Sandage} interpreted cosmology as the search for
two numbers: $ H_{0}$ and $q_{0}$.
It seemed too simple and clear: the main term
in the form of the Hubble parameter (HP) determined the expansion rate of
the Universe and a small correction due to the gravity of its matter content will slow down the expansion over time.
In 1970, Weinberg \cite{Winberg}  drew attention to the issue of extracting the value of constant spatial curvature $k$ and deceleration parameter $q$ from the observations,  without considering cosmological constant and/or scalar field. In 1976 Harisson \cite{Harrison} challenged Sandage's remark  and proved that  the third derivative of scale factor is of great importance for observational cosmology, in a universe with indeterminate (dust) matter density.
He considered a Universe containing the cosmological constant $\Lambda$
and non-relativistic matter. In this case, the Einstein equations reduce to the following
Friedmann equation(s)
 \begin{align}\label{h0}
&H^{2}+\frac{kc^{2}}{a^{2}}=\frac{8\pi G}{3}\rho+\frac{\Lambda}{3}\\
&\dot{H}+H^{2}=-\frac{4\pi G}{3}(\rho+3P)+\frac{\Lambda}{3}\label{h00}
\end{align}
For zero-pressure model, the equations (\ref{h0}) and (\ref{h00}) can be combined as
 \begin{align}\label{h1}
K=4 \pi G\rho-H^{2}(q+1)
\end{align}
Where $K=\frac{kc^{2}}{a^{2}}$, $\rho$ is the average mass density. The hubble $H$ and deceleration $q$  parameters are
 \begin{align}\label{h2}
H=\frac{\dot{a}}{a},\\ q=-\frac{\ddot{a}}{aH^{2}}
\end{align}
  The verification of the cosmological equation (\ref{h1}) requires the measurement of the quantities $K,H,\rho$ and $q$ \cite{Harrison}. In principle, these quantities $K,H,q$ can be determined, although in practice their precise determination is difficult\cite{Sandage}-\cite{Harrison}. If every thing could be known about the matter filling of universe, then the cosmological constant can be derived as
  \begin{align}\label{h23}
\Lambda=4\pi G- 3qH^{2}
\end{align}
Hence, for a universe with known amount of matter, the general relativity (GR) can be tested by measuring $H$ and $q$.  However, the average density $\rho$ can not be determined, even in principal \cite{Sandage}-\cite{Harrison}. Here, the third derivative of scale factor
  \begin{align}\label{h33}
Q=\frac{\dddot{a}}{aH^{3}}
\end{align}
is required to test the validity of equation (\ref{h1}).
Since the parameters $(k, \Lambda,\rho)$  can be obtained in terms of the first three derivatives of scale factor as
\begin{align}\label{h000}
K=H^{2}(Q-1)\\
\Lambda=H^{2}(Q-2q)\\
4\pi G\rho=H^{2}(Q+q)\label{h001}
\end{align}
So, the
resulting equations do not contain those parameters and can be expressed in
terms of the cosmological scalars as
\begin{align}\label{hh000}
X+2(q+Q)+qQ=0;
\end{align}
Where $X$ is the forth derivative of scale factor.
\begin{align}\label{hh0000}
X=\frac{\ddddot{a}}{aH^{4}}
\end{align}
The fourth order ODE (\ref{hh000}) is equivalent to the Friedmann equation (\ref{h1}) and has an
advantage that it appears as a constraint on directly measurable quantities.
In particular if $k = 0$ this
relation reduces to a third order ODE $Q=1$.\\
 Following the Harrison attempt, Hut \cite{Hut} clearly stated that without any presumption about  the matter content of the Universe, general relativity could not be tested by measuring any number of time derivatives of the scale factor. However, assuming a Universe filled with a non-interacting mixture of non-relativistic matter and radiation, the theory can be tested by measuring the first five derivatives of the scale factor\cite{Hut}.
 \begin{align}\label{hs00}
&k=\frac{1}{2}\Big(-X+4Q+qQ+2q-2\Big)H^{2}\\
&\Lambda=\frac{1}{4}\Big(-X+6Q+qQ+6q\Big)H^{2}\\
&\rho_{m}=\frac{1}{4 \pi G}\Big(-X+3Q+qQ+3q\Big)H^{2}\\
&\rho_{r}=\frac{3}{32 \pi G}\Big(X-2Q-qQ-2q\Big)H^{2}\label{hs00h}
\end{align}
By differentiating the equation (\ref{h1}) for the fourth time, finally, the parameters $(k, \Lambda,\rho_{m},\rho_{r})$ would be vanished and a relation between four parameters $q,Q,X,Y$ is obtained
 \begin{align}\label{hy00}
Y=7X+3Xq-6(Q+q)+Q^{2}-7Qq-3Qq^{2}+Qq^{2}
\end{align}
Where,
\begin{align}\label{hh000c}
Y=\frac{1}{H^{5}}\frac{1}{a}\frac{d^{5}(a)}{dt^{5}}
\end{align}
The condition (\ref{hy00}) provides a full test of the general relativity theory.
Hence, expressing the Friedmann equation in terms of the cosmographic parameters $H, q, Q, X, Y$ as higher derivatives of the scale factor,
 links the measurement of these parameters to a test
of GR or any of its modifications (leading to the different constraints). Cosmography was first
discussed by Weinberg \cite{Weinberg2 } and Visser \cite{Visse}, and has been extended by Capozziello \cite{Capozziello}-\cite{cap2} in a wide area of cosmological models. There is  also wide variety of studies that  implement this framework as a powerful model-independent approach to trace the history of the Universe \cite{Alam}-\cite{Zou}.
\\

Despite the assumption of the cosmological
constant in the above equations, the common idea at the time was just the Big Bang theory complemented with the inflation scenario, as the proper model of the Universe at least at first approximation.

However, the situation drastically
changed at the end of the last century.
Now, we know that the observations of the high redshift type Ia
supernovae and the surveys of the galactic clusters  \cite{Reiss}--\cite{Pope} reveal the accelerating expansion of Universe
and that the matter contribution is smaller than what was expected to be probably close to one. Also, the observations of Cosmic Microwave Background (CMB)
anisotropies indicate that the universe is flat and the total energy
density is very close to the critical one \cite{Spergel}.\\

Although the above observational methods are different, they are properly consistent with each other for the mass-energy content contributions of the universe. The data indicate that the universe at the present time is made up of $\%5$ normal mater, dark matter is estimated to be about $\%27$  and dark energy (DE) is the dominant component that occupies about $\%68$ of the total energy content.

There are prominent candidates for DE such as the cosmological
constant \cite{Sahni, Weinberg}, a dynamically evolving scalar field ( like quintessence) \cite{Caldwell, Zlatev} or phantom (field with negative energy) \cite{Caldwell2} that explain the cosmic accelerating expansion. Meanwhile, the accelerating
expansion of universe can also be obtained through
modified gravity \cite{Zhu}, brane cosmology and so on \cite{Zhu1}--\cite{nojiri}.

The archetypal example of scalar field is chameleon field which has
been suggested by \cite{Khoury}--\cite{Khourym}.
The cosmological value of such a field evolves over Hubble time-scales and could potentially cause the late-time acceleration of our Universe \cite{Brax2}. The Chameleon mechanism is nowadays deeply investigated in many of
possible "shapes" (i.e. varied potentials and theoretical
backgrounds), on all scales (from cosmological to astrophysical) and
by much more reliable (mainly numerical) methods\cite{Brax1}-\cite{Haidar}.\\
In this paper we want to express the chameleon parameters in terms of directly measurable
cosmological scalars constructed out of higher derivatives of the scale factor (cosmographic parameters). It enable us to
construct model-independent kinematics of the chameleon cosmology. Similar to the way  that Harisson \cite{Harrison} and Hut \cite{Hut} had done to test the general relativity, we want to find the algebraic relations which can be used to derive the free parameters in terms of the cosmographic ones and those are between latter parameters by assuming that the chameleon cosmology is hold.
We consider an exponential potential $V=M^{4}Exp({\frac{\alpha\phi}{M_{pl}}})$ where $M$ and $\alpha$ are the  mass scale and coupling constant, respectively. In section. 2, we adopt the relations that have been previously found to test the model when the system dynamically approaches to its critical points. The summary and conclusion are presented in section 3.

\section{The Model}

We begin with the action of chameleon gravity given by,
\begin{eqnarray}\label{action}
S=\int[\frac{M_{Pl}^2}{16\pi}{\cal R}-\frac{1}{2}\phi_{,\mu}\phi^{,\mu}+V(\phi)]\sqrt{-g}dx^{4}\nonumber\\+\int {\cal L}_m(\psi^{(i)}, g_{\mu\nu}^{(i)})dx^{4},
\end{eqnarray}
where the matter fields $\psi^{(i)}$ are coupled to scalar field $\phi$ by the definition $g_{\mu\nu}^{(i)}\equiv e^{2\beta_i\phi/M_{Pl}}g_{\mu\nu}$.
The $\beta_{i}$ are dimensionless coupling constants, one for each type of matter. In the following, we assume a single matter energy density component $\rho_m$ with coupling $\beta$ \cite{Khourym}. Assuming that the universe is filled with cold dark matter, i.e. $\gamma=0$, the variation of action (\ref{action})  with respect to the metric tensor components in a spatially flat FRW  cosmology yields the field equations,
\begin{eqnarray}\label{fried1}
3H^{2}M_{pl}^{2}=V_{eff}(\phi)+\frac{1}{2}\dot{\phi}^{2},
\end{eqnarray}
\begin{eqnarray}\label{frir2}
2\dot{H}M_{pl}^{2}=-\dot{\phi}^{2}-\rho_{m}e^{\frac{\beta}{M_{pl}}\phi}
\end{eqnarray}
Where, the chameleon effective potential is defined by,
\begin{eqnarray}\label{veff0}
V_{eff}(\phi)=V(\phi)+\rho_{m}e^{\frac{\beta}{M_{pl}}\phi},
\end{eqnarray}
 providing the wave
equation of chameleon scalar field $\phi$.
\begin{eqnarray}\label{phiequation}
\ddot{\phi}+3H\dot{\phi}=-\frac{dV_{eff(\phi)}}{d\phi},
\end{eqnarray}

One can also easily find the mass associated with  field $\phi$:
\begin{eqnarray}
m_{ch}^{2}=\frac{d^{2}}{d\phi^{2}}V_{eff}(\phi).
\end{eqnarray}
For $\gamma=0$, if $\beta$ in the second term of $V_{eff}(\phi)$ is positive, the effective potential monotonically decreases to a minimum at a finite field value $\phi=\phi_{min}$, where $\frac{d}{d\phi}V_{eff}(\phi)|_{\phi=\phi_{min}}=0$, and $ m_{ch}=m_{ch_{min}}$. From equation (\ref{phiequation})
\begin{eqnarray}\label{phiequation2}
\ddot{\phi}_{min}=-3H\dot{\phi}_{min}
\end{eqnarray}
Note that $m_{ch_{min}}$ in the above equations is the inverse of the characteristic range of
 chameleon force in a given medium. By differentiating both sides of the equation  (\ref{frir2}) with respect to time

\begin{eqnarray}\label{frr2}
2\ddot{H}M_{pl}^{2}=-2\dot{\phi}\ddot{\phi}-\dot{\rho_{m}}e^{\frac{\beta}{M_{pl}}\phi}-\frac{\beta}{M_{pl}}\rho_{m}e^{\frac{\beta}{M_{pl}}\phi}\dot{\phi}\nonumber\\
\end{eqnarray}
The equation (\ref{frr2}) can be rewritten as
\begin{eqnarray}\label{fr2}
2\ddot{H}M_{pl}^{2}=-2\dot{\phi}\ddot{\phi}-3H\dot{\phi}^{2}-6H\dot{H}M_{pl}^{2}+2\beta M_{pl}\dot{\phi}\dot{H}+\frac{\beta}{M_{pl}}\dot{\phi}^{3}\nonumber\\
\end{eqnarray}

From equations (\ref{fried1}) and \ref{frir2} we can also obtain
\begin{align}\label{vv}
&\rho_{m}e^{\frac{\beta}{M_{pl}}\phi}=-2M_{pl}^{2}\dot{H}-\dot{\phi}^{2}\\
&V(\phi)=(2\dot{H}+3H^{2})M_{p}^{2}+\frac{1}{2}\dot{\phi}^{2}\label{vv1}
\end{align}
Using equations (\ref{vv}) and (\ref{vv1}) and by considering exponential potential $V=V_{0}e^{\frac{\alpha}{M_{pl}}\phi}$ where dimensionless
constant $\alpha$   characterizing the slope of potential, one can obtain
\begin{align}\label{ph0}
\frac{dV_{eff}}{d\phi}=2(\alpha-\beta)M_{pl}\dot{H}+\frac{(\alpha-2\beta)}{2M_{pl}}\dot{\phi}^{2}+3\alpha M_{pl}H^{2}
\end{align}
Also, the mass of chameleon field is expressed by
\begin{align}\label{p}
m_{ch}^{2}=\frac{d^{2}V_{eff}}{d\phi^{2}}=2(\alpha^{2}-\beta^{2})\dot{H}+\frac{(\alpha^{2}-2\beta^{2})}{2M_{pl}^{2}}\dot{\phi}^{2}+3\alpha^{2} H^{2}
\end{align}
So, using equations (\ref{phiequation}) and (\ref{ph0}) the second time derivative of scalar $\phi$ would be
\begin{eqnarray}\label{phiequation2}
\ddot{\phi}=-3H\dot{\phi}-2(\alpha-\beta)M_{pl}\dot{H}-\frac{(\alpha-2\beta)}{2M_{pl}}\dot{\phi}^{2}-3\alpha M_{pl}H^{2}\nonumber,\\
\end{eqnarray}

By combining the equations (\ref{phiequation2})and (\ref{fr2}), the second derivative of Hubble parameter gives

\begin{align}\label{fre}
&2\frac{\ddot{H}}{H^{3}}=-6\frac{\dot{H}}{H^{2}}+(\alpha-\beta)(\frac{\dot{\phi}}{M_{pl}H})^{3}+3(\frac{\dot{\phi}}{M_{pl}H})^{2}\nonumber
\\&+\Big((2\beta-4\alpha)\frac{\dot{H}}{H^{2}}+2\alpha\Big)(\frac{\dot{\phi}}{M_{pl}H})\nonumber\\
\end{align}
Here, we rewrite the cosmographic parameters as follows
\begin{align}\label{q0}
&q=-1-\frac{\dot{H}}{H^{2}}\\
&Q=\frac{\ddot{H}}{H^{3}}-3q-2\\
&X=\frac{\dddot{H}}{H^{4}}+4Q+3q(q+4)+6\\
&Y=\frac{\ddddot{H}}{H^{5}}+5X-10Q(q+2)-30q(q+2)-24\label{y0}
\end{align}

Here for more simplicity, we also define the following dimensionless parameters
\begin{eqnarray}\label{new variable}
x=\frac{\phi\dot{}}{H M_{pl}},y=\frac{\rho_{m}e^{\frac{\beta\phi}{M_{pl}}}}{H^{2}M^{2}_{pl}},z=\frac{V}{H^{2}M^{2}_{pl}}
\end{eqnarray}
Where equation (\ref{fried1}) puts a constraint on the variables as
\begin{eqnarray}\label{new variables}
1=\frac{x^2}{6}+\frac{y}{3}+\frac{z}{3}
\end{eqnarray}
Equation (\ref{fre}) can be rewritten in terms of the cosmographic parameters $(q,Q)$ and the model parameters $(x,\alpha,\beta)$ as
\begin{align}\label{co}
&2\Big(Q-1\Big)=(\alpha-\beta)x^{3}+3x^{2}+\Big((4\alpha-2\beta)(1+q)+2\alpha\Big)x\nonumber\\
\end{align}
 This is a cubic polynomial in terms of $x$, as $ax^{3}+bx^{2}+cx+d=0$, where\\
 $\left\{
\begin{array}{ll}
a=(\alpha-\beta)\\
b=3\\
c=\Big((4\alpha-2\beta)(1+q)+2\alpha\Big)\\
d=2\Big(1-Q\Big)\\
\end{array}
\right.
$\\

Solving the equation (\ref{fre}), therefore, gives $x$ in terms of the cosmographic  $(q,Q)$ and constant  $(\alpha,\beta)$ parameters.

For a simple case where $Q=1$, one can find the following three solutions\\
 $\left\{
\begin{array}{ll}
x=0\\
x=\frac{-3+(9+16\alpha^2q-24\alpha\beta q-8\alpha^2+8\beta^2+8\beta^2q)^{\frac{1}{2}}}{2(\alpha-\beta)}\\
x=-\frac{3+(9+16\alpha^2q-24\alpha\beta q-8\alpha^2+8\beta^2+8\beta^2q)^{\frac{1}{2}}}{2(\alpha-\beta)}\\
\end{array}
\right.
$\\
It is a striking and slightly puzzling fact that
almost all current cosmological observations can be summarized by a simple
statement: The jerk of the Universe is equal to one” $(Q=1).$ \cite{Maciej},\cite{Alam},\cite{Kun}. Also Visser\cite{Visse}, have investigated in some details the jerk condition $(Q=1)$ .
 From equations $(\ref{vv})$ and $(\ref{vv1})$,  the variables $(y,z)$  can also be specified in terms of $x,q$;
\begin{align}\label{rm}
&y=-x^{2}+2(1+q)\\
&z=\frac{x^{2}}{2}+(1-2q)\label{rm2}
\end{align}
If the values of $(\alpha,\beta)$ are known, consequently, all model variables would be reconstruct in terms of cosmographic parameters $(q,Q)$.
 In addition, the equation (\ref{p}) gives the chameleon mass $m_{ch}$ in terms of $(H,q,Q,\alpha,\beta)$
\begin{align}\label{mm}
\frac{m_{ch}^{2}}{H^{2}}=2(\beta^{2}-\alpha^{2})(1+q)+\frac{(\alpha^{2}-2\beta^{2})}{2}x^{2}+3\alpha^{2}
\end{align}
Equations (\ref{co}) and (\ref{mm}) imply the possibility of determining the chameleon mass in terms of coupling parameters $(\alpha,\beta)$ and first three derivatives of scale factor $(H,q,Q)$. Note that l.h.s of equation (\ref{xx}) is positive, hence $x^{2}<2(1+q)$. Exerting this condition on equation (\ref{mm}) gives an upper bound for chameleon mass as $m_{ch}^{2}<\alpha^{2}(2-q)H^{2}$. If the parameters $(\alpha,\beta)$ are known, chameleon mechanism can be interpenetrated by only the first three derivatives of scale factor $(H,q,Q)$. As it was pointed out by the original chameleon article that \cite{Khourym}, in harmony with string theory, $\beta$ must be of the order unity; so that we got $\beta=1$. Furthermore, we get $\alpha$ of the order of unity with negative sign (because the potential $V(\phi)$) is assumed to be of the runaway
form for monotonically decreasing function of $\phi$), then the solution of equation (\ref{co}) gives
\begin{align}\label{xx}
x=\frac{1}{2}+\frac{1}{6}A^{\frac{1}{3}}-(\frac{13}{2}+6q)A^{\frac{-1}{3}}
\end{align}

Where,\\ $A=6\Big(1830+5292q+4941q^2+1296q^3+486Q+972qQ+324Q^2\Big)^{\frac{1}{2}}-81-162q-108Q$\\
Finally, using equations (\ref{xx}), (\ref{rm}) and (\ref{rm2}), the chameleon gravity is expressed on the way of time derivatives of scale factor measuring the  $(H,q,Q)$ parameters.

However, if nothing can be proposed about $(\alpha,\beta)$, the chameleon parameters will not be determined by considering the first three derivatives of scale factor, as its forth and fifth derivatives  must also be measured.
By differentiating both sides of the equation (\ref{frir2}) with respect to time, the third time derivative of Hubble parameter would be obtained as

\begin{align}\label{fr}
2\dddot{H}M_{pl}^{2}&=-2\dot{\phi}\dddot{\phi}-2\ddot{\phi}^{2}-3\dot{H}\dot{\phi}^{2}-6H\dot{\phi}\ddot{\phi}-6H\ddot{H}M_{pl}^{2}
\nonumber\\&-6\dot{H}^{2}M_{pl}^{2}+2\beta M_{pl}\ddot{\phi}\dot{H}+2\beta M_{pl}\dot{\phi}\ddot{H}+3\frac{\beta}{M_{pl}}\ddot{\phi}\dot{\phi}^{2}
\end{align}
Differentiating equation (\ref{phiequation}) also yields
\begin{align}\label{ph}
\dddot{\phi}=-3H\ddot{\phi}-3\dot{H}\dot{\phi}-2(\alpha-\beta)M_{p}\ddot{H}-\frac{(\alpha-2\beta)}{M_{p}}\dot{\phi}\ddot{\phi}
-6\alpha M_{pl}H\dot{H}
\end{align}
Their composition is expressed by
\begin{align}\label{ph1}
&(X-4Q)=\\ \nonumber
&-6-2\beta^2-\alpha^2-3\alpha\beta+3\alpha\beta q+4\alpha^{2}q
-4\alpha^2q^2-2\beta^2q^2\\ \nonumber
&-3q+6\alpha\beta q^2-4\beta^{2}q+x(-9\alpha-15\beta+18\alpha q-15\beta q)\\ \nonumber
&+x^{2}(3\beta^2-\frac{9}{2}\alpha\beta q
-\frac{3}{2}q-15+3\beta^2q)+(\frac{15}{2}\beta-\frac{9}{2}\alpha) x^3\\ \nonumber
&+x^{4}(\frac{1}{4}\alpha^2+\frac{3}{4}\alpha\beta-\beta^{2})\\ \nonumber
\end{align}

Corresponding to what was previously achieved for the equation (\ref{ph1}), by differentiating both sides of equations (\ref{fr}) and (\ref{ph}) and using equations (\ref{q0}) to (\ref{y0}), we take
\begin{align}\label{ph2}
&(Y-5X)=\\ \nonumber
&19+29q-\frac{157}{2}\beta x^3-16 x \alpha q^2+17 x \beta q^2+q x^3 \alpha-7 q \beta x^3 \\ \nonumber
&+16x\alpha^3 q+5 \alpha x \beta^2-16 \alpha^3 q^2 x+6 \alpha^3 q x^3-16 x \beta^3 q-7 \beta x \alpha^2\\ \nonumber
&-8 \beta^3 q^2 x+9 \beta^3 q x^3+\frac{5}{4} x^5 \alpha \beta^2+\frac{7}{4} x^5 \beta \alpha^2-4 x \alpha^3-3 x^3 \alpha^3\\ \nonumber
&+9 \beta^3 x^3-\frac{1}{2} x^5 \alpha^3-\frac{5}{2} x^5 \beta^3+\frac{3}{4} x^5 \alpha-\frac{3}{4} \beta x^5+\frac{9}{4} x^4-8 x \beta^3\\ \nonumber
&+114 x^2+9 \alpha^2-42 \alpha \beta q+\frac{69}{2} x^2 q-\frac{21}{4} x^4 \alpha^2+\frac{45}{2} \beta^2 x^4\\ \nonumber
&+42 \beta^2 q^2+6 q^2-66 x^2 \beta^2 q+12 x^2 \alpha^2 q-84 \alpha q^2 \beta-\frac{21}{2} x^4 \alpha \beta\\ \nonumber
&+14 \beta q^2 x \alpha^2-\frac{21}{2} \beta q x^3 \alpha^2+63 \alpha q x^2 \beta+5 \alpha x \beta^2 q\\ \nonumber
&-\frac{21}{2} \alpha q \beta^2 x^3+7 \beta x \alpha^2 q+177 x \beta q-114 x \alpha q+61 x \alpha\\ \nonumber
&+42 \beta^2-36 \alpha^2 q+42 \alpha \beta-6 x^2 \alpha^2+84 \beta^2 q-66 x^2 \beta^2\\ \nonumber
&+36 \alpha^2 q^2+10 \alpha q^2 x \beta^2+160 x \beta+32 x^3 \alpha
\end{align}

 \subsection{Distinguish between "cosmographic test of the model"and "reconstructing the model in terms of cosmographic parameter"}
 A subtle and important point that must be point out is that we must distinguish between two conceptions "reconstruction of the model in terms of cosmographic parameters" and "test of the model using cosmographic parameters". In considerations such as the former, in fact, it is assumed that the model is theoretically valid. Based on this, the parameters of the model are reconstructed in terms of the measurable (observable) cosmographic parameters In latter cases, however, the aim is to find the relationships between measurable cosmographic parameters to test the theory of the model. On the other hand, it also tries to eliminate the parameters of the model to find the algebraic relation between model independent cosmographic parameters. A  condition should be stated to say that the model is fully tested. While the model parameters are reconstructed in both of these considerations, more cosmographic parameters are needed to test the theory. For example, for $\Lambda CDM$ model with free parameters $\Lambda$, zero-pressure matter $\rho$, and $K$ , all model parameters can be reconstructed according to the first three derivatives of the scale factor $(H,q,Q)$ as equations (\ref{h000}) to (\ref{h001}) . However, a full test of the model can be done by equation (\ref{hh000}) where includes the forth derivative of the scale factor $X$.
Another example, that has been investigated in some details by Hut, is a universe filled with a combined density of matter and radiation. In this case the free parameters of the model are $\Lambda,\rho_{m},\rho_{r},K$ constructed by cosmographic parameters $(H,q,Q,X)$ as equations (\ref{hs00})-(\ref{hs00h}) . Nevertheless, the full test of the model is codified in the equation (\ref{hy00}), where includes the fifth derivative of the scale factor $Y$.
In chameleon model, the relations were obtained between the four (in principle free) parameters of the model and the first five cosmographic parameters, which can be written in terms of the Hubble parameter and its first four derivatives. The four free parameters include two new variables $(x,y)$ and two free parameters $(\alpha,\beta)$.
  (Note that the variable $z$ can be obtained in terms of $(x,y$) from constraint Eq. (\ref{new variables}).)\\
  If such cosmographic parameters are known, then it is possible to present  algebraic relations that can be used to derive the free parameters in terms of the cosmographic ones.\\
  Depending on whether the parameters $(\alpha,\beta)$ are known or not, we can classify our analysis and discussion in two cases, as follows \\
  \textbf{1}-$(\alpha,\beta)$ \textbf{is known} : In this case, the only free variables or parameters of the model are $(x,y)$. Hence, using equation (\ref{co}),
the variable $x$ can be reconstructed in terms of $(\alpha,\beta,q,Q)$. Then from equations (\ref{co}) and (\ref{rm}), the variables $(y,z)$ can also be reconstructed in terms of $(x, q)$.\\
Assuming that  $(\alpha,\beta)$ are known, hence, $(q,Q)$ are the only cosmographic parameters that need to be determined to reconstruct the chameleon parameters.\\
To test the model, however, only the  $(q,Q)$ parameters are not sufficient, as the forth derivative of scale factor $X$ is required to find a relation between  $(q,Q,X)$ quantities. In this respect, we should figure out the variable $x$ from equation (\ref{co}); so that by substituting it in the equation (\ref{ph1}), we can obtain a condition for providing a full test of the model (note that the condition includes the known parameters $(\alpha,\beta)$). \\
\textbf{2}-$(\alpha,\beta)$ \textbf{is unknown} : This case is appeared to be more realistic than pervious ones. If nothing could
be proposed about the parameters $(\alpha, \beta)$, chameleon gravity could be reconstructed by measuring the first five
derivatives of the scale factor and could also be tested by measuring its first six
derivatives.
There are four free parameter $(\alpha,\beta,x,y)$ by considering undetermined $(\alpha, \beta)$. The equations (\ref{co}), (\ref{rm}), (\ref{ph1}) and (\ref{ph2}) are sufficient to reconstruct the chameleon free parameters in terms of the cosmographic ones $(q,Q,X,Y)$. An interesting feature is that, in this case
 the parameters $(\alpha, \beta)$ can also be determined in terms of the cosmographic parameters. However, to test the chameleon gravity we require This requires an additional cosmological
scalar  $M=\frac{1}{H^{6}}\frac{1}{a}\frac{d^{6}(a)}{dt^{6}}$ to find a relation just between cosmographic parameters, which we call it as the full test of chameleon theory.
Equations (\ref{co}), (\ref{ph1}) and (\ref{ph2})  with seven variables $(x,\alpha,\beta,q,Q,X,Y)$ indicating that, if four cosmographic parameters $(q,Q,X,Y)$ are known, the chameleon parameters $(\alpha,\beta)$ and variable $x$ will be determined immediately. Consequently, the chameleon free parameters can be reconstructed on cosmological scales.\\
For more explanation, we want to reconstructing the chameleon free parameters $(\alpha,\beta)$ and current values of new variables $(x_{0},y_{0},z_{0})$ of the model in terms of determined cosmographic parameters. Here we consider the values that best fitted  by observations \cite{Tegmark} \\
$q_{0}=-.588,Q_{0}=1,X_{0}=-.238,Y_{0}=2.846$\\
From equation (\ref{co}), there are three solutions for the variable $x$ when $Q_{0}=1$, as $x=0$ is a simple one.\\
In this case, according to the equation (\ref{rm}), $y_{0}=2(1+q_{0})=0.824$. By subsisting these best fit values of cosmographic parameters in the equations (\ref{ph1}) and (\ref{ph2}), two coupled equations obtain in terms of $(\alpha,\beta)$ as follows\\

    $\left\{
\begin{array}{ll}
-4.734976\alpha^2-.339488\beta^2-2.689536\alpha\beta-4.236=0\\
42.614784\alpha^2+7.129248\beta^2+37.653504\alpha\beta+4.022464 =0\\
\end{array}
\right.
$\\

Solving these equations gives\\
$\left\{
\begin{array}{ll}
(\alpha\simeq-2.780,\beta\simeq10.01)\\
(\alpha\simeq2.780,\beta\simeq-10.01)\\
\end{array}
\right.
$\\
Hence, the four free parameters of the chameleon model is reconstructed.
Furthermore, the author of \cite{Cap}, finds the best values for cosmographic parameters as
\begin{align}
q_{0}=-.64,Q_{0}=1.02,X_{0}=-.39,Y_{0}=4.05\nonumber
\end{align}

Reconstructing the chameleon parameters for this case gives\\
\begin{align}
x_{0}=0,y_{0}=0.74,\ \ (\alpha,\beta)\simeq( \pm2.598, \mp 11.25)\nonumber
\end{align}\\
In the above examples, the current values of model parameters were determined, since the current value of cosmographic parameters are available. However it is possible to reconstruct the model in terms of cosmographic parameter at any epoch of the universe
In the following section we present the relation between cosmographic parameters in terms of free parameters of the model at crucial epochs of the universe
\section{Cosmography at different epochs of the Universe}

As it goes to describe the behaviour of a dynamical system, we implement this approach to trace the dynamics of the universe in the presence of chameleon field in deferent epochs of its evolution.

 We are interested to quantify the chameleon parameters such as chameleon mass $m_{ch}$  at the critical points of the system,  points that represent
all important epochs in the evolution of the universe. The dynamics of the universe in chameleon gravity can be simplified by introducing the
following dimensionless variables,
\begin{eqnarray}\label{new variable}
\zeta_{1}=\frac{\phi\dot{}}{\sqrt{6}H M_{pl}},\zeta_{2}=\frac{\rho_{m}e^{\frac{\beta\phi}{M_{pl}}}}{3H^{2}M^{2}_{pl}},\zeta_{3}=\frac{V}{3H^{2}M^{2}_{pl}}
\end{eqnarray}
 Then using
equations (\ref{fried1})-(\ref{phiequation}), the evolution equations of these
variables become,
\begin{eqnarray}
\zeta_{1}^{'}&=&-3\zeta_{1}+\frac{3}{2}\zeta_{2}\zeta_{1}
+3\zeta_{1}^{3}-\frac{\sqrt{6}}{2}\beta \zeta_{2}-\frac{\sqrt{6}}{2}\alpha\zeta_{3} ,\label{x1}\\
\zeta_{2}^{'}&=&-3\zeta_{2}+3\zeta_{2}^{2}+\sqrt{6}\beta \zeta_{1} \zeta_{2}
+6\zeta_{1}^{2}\zeta_{2},\label{y1}\\
\zeta_{3}^{'}&=&\zeta_{3}(-\sqrt{6}\alpha\zeta_{1}+3\zeta_{2}+6\zeta_{2}^{2})\label{z1}
\end{eqnarray}

Where prime indicates from now on differentiation with respect to $N =ln a$. Note that $\zeta_{1}=\frac{x}{\sqrt{6}},\zeta_{2}=\frac{y}{3},\zeta_{3}=\frac{z}{3}$. The Friedmann  equation (\ref{fried1}) also becomes
\begin{eqnarray}\label{const}
\zeta_{1}^2+\zeta_{2}+\zeta_{3}=1
\end{eqnarray}

Using the constraint (\ref{const}), the equations (\ref{x1})-(\ref{z1}) are converted to
\begin{align}
&\zeta_{1}^{'}=\zeta_{1}(-3+\frac{3}{2}\zeta_{2}
+3\zeta_{1}^{2})-\frac{\sqrt{6}}{2}\Big(\beta-\alpha) \zeta_{2}+\alpha(1-\zeta_{1}^{2})\Big) ,\nonumber \label{x2}\\\\
&\zeta_{2}^{'}=-3\zeta_{2}+3\zeta_{2}^{2}+\sqrt{6}\beta \zeta_{1} \zeta_{2}
+6\zeta_{1}^{2}\zeta_{2},\nonumber\label{y2}\\
\end{align}
In terms of the new dynamical variable, we also have
\begin{eqnarray}\label{dec}
q=-\Big(1+\frac{\dot{H}}{H^{2}}\Big)=-1+\frac{3\zeta_{2}}{2}+3\zeta_{1}^{2}
\end{eqnarray}
One can also rewrite the equations (\ref{co}) and (\ref{mm}) as
\begin{align}\label{co2}
&2\Big(Q-1\Big)=36\sqrt{6}(\alpha-\beta)\zeta_{1}^{3}+18\zeta_{1}^{2}\nonumber
\\&+\sqrt{6}\Big((4\alpha-2\beta)(1+q)+2\alpha\Big)\zeta_{1}\nonumber\\
\end{align}
\begin{align}\label{mm2}
\frac{m_{ch}^{2}}{H^{2}}=2(\beta^{2}-\alpha^{2})(1+q)+3(\alpha^{2}-2\beta^{2})\zeta_{1}^{2}+3\alpha^{2}
\end{align}
It is more convenient to investigate the properties of the dynamical system, namely  Eqs.(\ref{x2}) and (\ref{y2})
 rather than Eqs.(\ref{x1})-(\ref{z1}). We obtain the fixed points (critical points) and study
the stability of these steady states that are always exact constant solutions in the
context of autonomous dynamical systems. Those are often the extreme points of
the orbits and therefore describe the asymptotic behavior. In the following we find fixed points by  solving $\frac{d\zeta_{1}}{dN}=0$ and $\frac{d\zeta_{2}}{dN}=0$ simultaneously. Two eigenvalues $\lambda_{i} (i=1,2)$ are obtained by substituting linear perturbations $\zeta_{1}'\rightarrow \zeta_{1}'+\delta \zeta_{1}'$, $\zeta_{2}'\rightarrow \zeta_{2}'+\delta \zeta_{2}'$ about the critical points into the two independent equations (\ref{x2}) and (\ref{y2}), to the first order of  perturbations. Stability requires that the real part  of all eigenvalues  be negative. There are also  five fixed points which some of them explicitly depend on  $\beta$ and $\alpha$, as shown in Table 1.
\begin{table}
\caption{\label{tmodel} critical points}
\begin{tabular}{cccccc}
points  &  $A_{+}$  &$A_{-}$ \ & $B$ \ & C \ & D  \\
\hline 
\hline
$\zeta_{1}$  &1 & -1 &$ -\frac{\sqrt{6}\alpha}{6} $& $-\frac{\sqrt{6}}{3}\beta $
& $\frac{\sqrt{6}}{2\beta-2\alpha} $\\
\hline
$\zeta_{2}$ & 0 & 0 & 0 & $1-\frac{2\beta^{2}}{3}
$&$\frac{-\beta\alpha+\alpha^{2}-3}{(-\beta+\alpha)^{2}} $ \\
\hline 
\hline\end{tabular}
\end{table}

{\bf Critical points, $ A_{\pm}$}, corresponding to two {\em kinetic-dominated} solutions.
These are equivalent to the stiff-fluid-dominated evolution with $a=t^{\frac{1}{3}} $, irrespective of the nature of  the potential. The kinetic-dominated solution for $A_{+}$ has two eigenvalues, $\lambda_+=3+\beta\sqrt{6},\lambda_-=6+\sqrt{6} \alpha$,
 \begin{figure}[t]
\includegraphics[scale=.45]{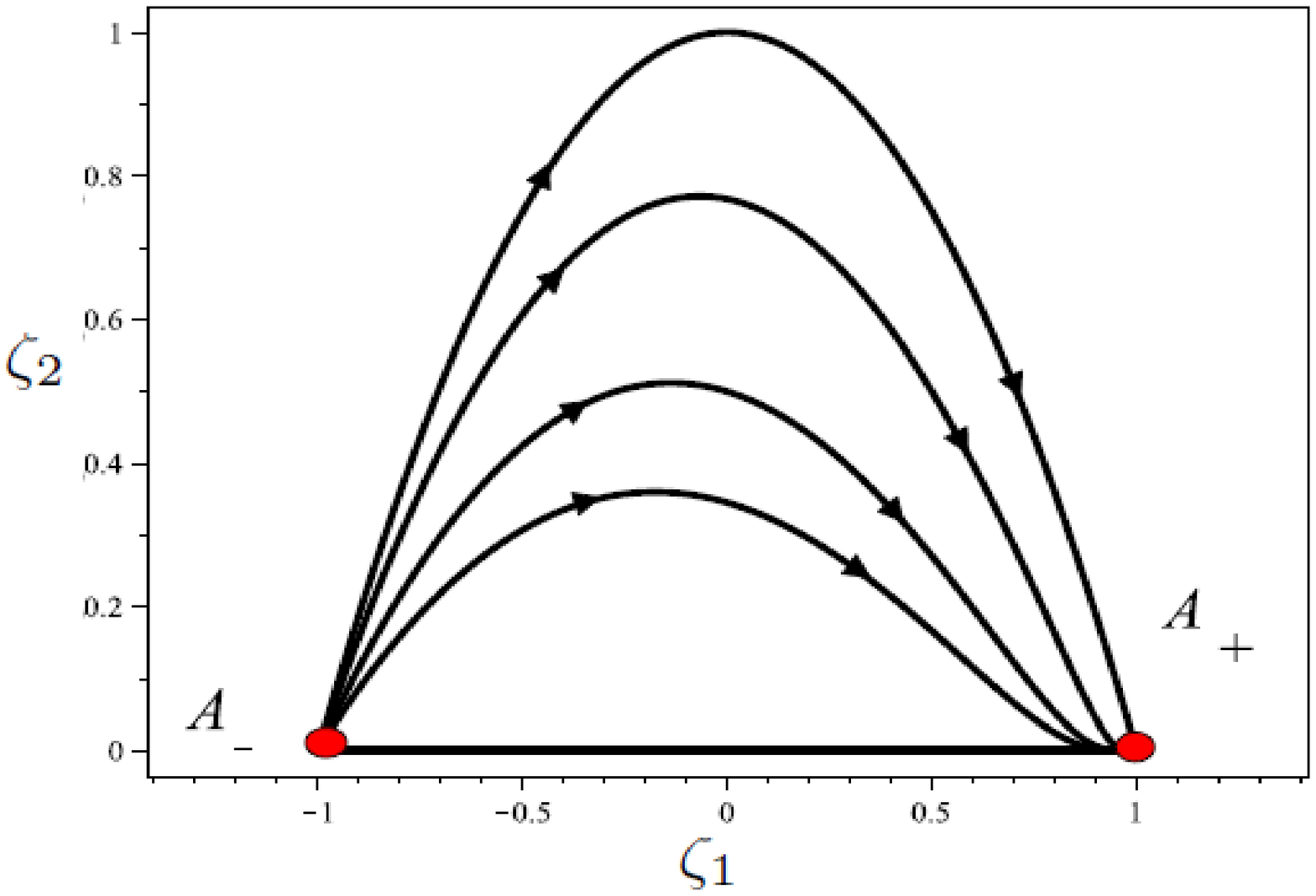}\hspace{0.1 cm}\\
Fig. 1:  The phase space for $\alpha=-3$ and $\beta=-4$ . The late time attractor is scaling solution with $\zeta_{1}=1,\zeta_{2}=0$ \\
\end{figure}
\begin{figure}[t]\includegraphics[scale=.45]{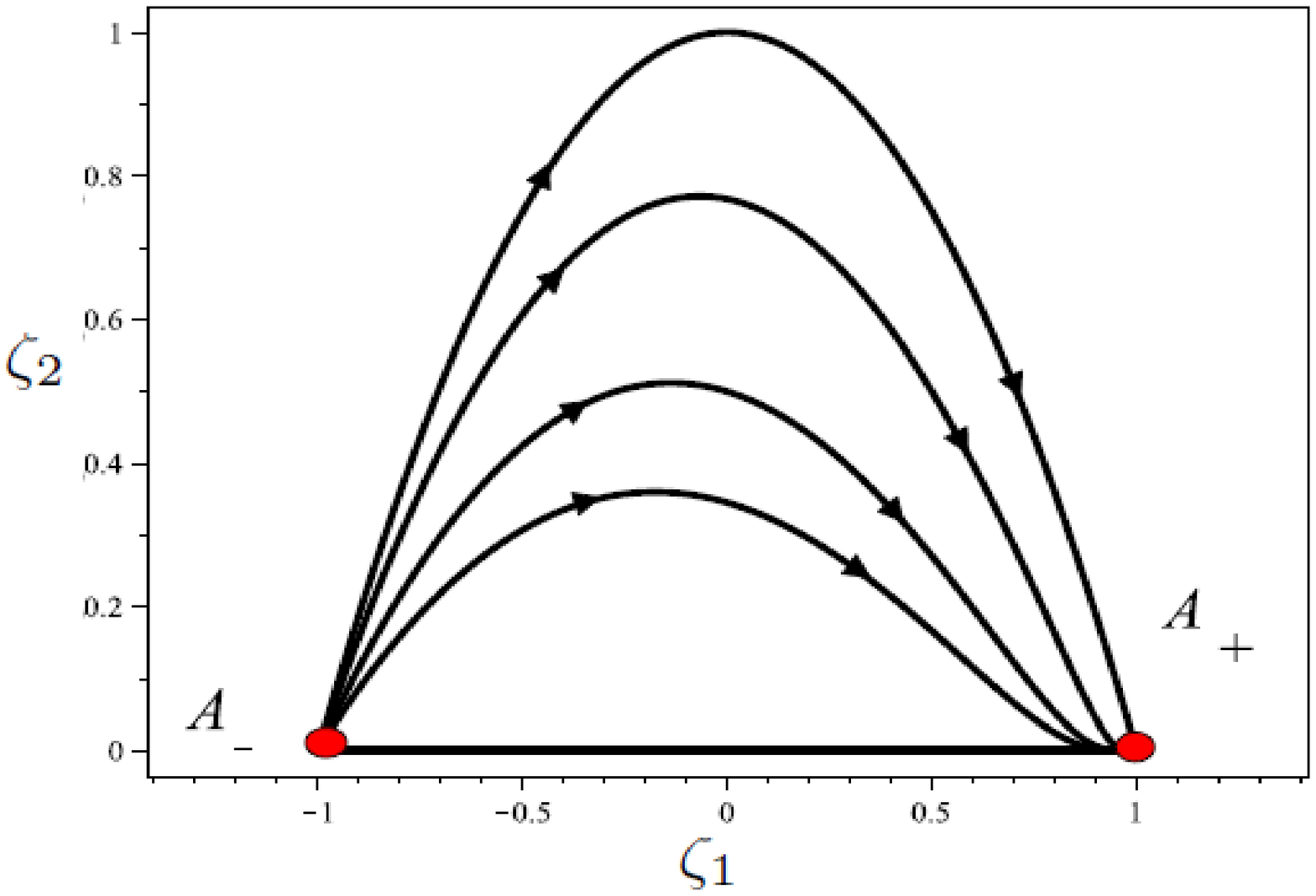}\hspace{0.1 cm}\\
Fig. 2:  The phase space for $\alpha=3$ and $\beta=4$ . The late time attractor is scaling solution with $\zeta_{1}=-1,\zeta_{2}=0$ \\
\end{figure}
 and is stable  for $\beta<\frac{-\sqrt{6}}{2}$,$\alpha<-\sqrt{6}$. The solution $A_{-}$ also has  two eigenvalues
$\lambda_+=3-\beta\sqrt{6}$, and $\lambda_-=6-\sqrt{6} \alpha$, stabilized  for $\beta>\frac{\sqrt{6}}{2}$ and $\alpha>\sqrt{6}$. The phase space of the system, in which these critical points are stable, has been shown in Figs. (1) and (2). Also, from equations \ref{dec} and (\ref{co}) to (\ref{ph2}),
the values of the cosmographic parameters at these points are as follows:\\
$ \left\{
   \begin{array}{ll}
   q_{c}=2\\
   Q_{c}=10\\
   X_{c}=-80\\
   Y_{c}=880\\
     \end{array}
 \right.
$\\
  where both  $V(\phi)$ and effective  $V_{eff}$ potentials are zero. So, the chameleon mass $m_{ch}$ at those points is zero, as confirmed by equation (\ref{mm2}).

{\bf Critical point, $B$,} corresponding to a {\em potential-kinetic-scaling} solution.
This solution exists for all kinds of  potentials, and has two eigenvalues depending on the slope of the potential and coupling constant $\beta$: $\lambda_+=-3+\frac{\alpha^2}{2} \,, \qquad \lambda_-=-\beta\alpha+\alpha^2-3.$\\ As, the solution
is stable  for\\
$ \left\{
   \begin{array}{ll}
    \beta<\frac{-3+\alpha^{2}}{\alpha}, -\sqrt{6}<\alpha<0\\  \beta>\frac{-3+\alpha^{2}}{\alpha}, 0<\alpha<\sqrt{6}  \hbox{$\ \ \ \ \ \ \ \  \ \ \ \ \ \ \ \ $}
   \end{array}
 \right.
$\\
which means that the potential-kinetic-dominated solution is stable for a sufficiently flat potential ($\alpha^{2}<6$). The phase space of the system  has been shown in Fig.(3), in which the critical point is stable.
The set of the values of the cosmographic parameters at this point is as follows
$ \left\{
   \begin{array}{ll}
   q_{c}=-1+\frac{1}{2}\alpha^{2}\\
   Q_{c}=\frac{1}{2}\alpha^{4}-\frac{3}{2}\alpha^{2}+1\\
  X_{c}=-\frac{3}{4}\alpha^{6}+\frac{11}{4}\alpha^{4}-3\alpha^{2}+1\\
   Y_{c}=\frac{3}{2}\alpha^{8}-\frac{25}{4}\alpha^{6}+\frac{35}{4}\alpha^{4}-5\alpha^{2}+1\\
      \end{array}
 \right.
$\\
The chameleon mass $m_{ch}$ would also be
 \begin{eqnarray}\label{mp0}
\frac{m_{ch}^{2}}{3H^{2}}=\alpha^{2}(1-\frac{\alpha^{2}}{6})
\end{eqnarray}
indicating that the stability condition $(\alpha^{2}<6)$ leads to an upper bound for chameleon mass as $m_{ch}^{2}<\frac{9H^{2}}{2}$. For $\alpha=\pm\sqrt{2}$ at this point,  corresponding to the $deceleration-acceleration$ phase of the universe,  $m_{ch}=2H$ and all  cosmographic parameters are zero.
 For $\alpha \rightarrow 0$ at this point, in addition, the potential tends to $V(\phi)\rightarrow M^{4}$ and the cosmographic parameters would be $ \{q_{c}=-1,Q_{c}=X_{c}=Y_{c}=1\}$ , representing the cosmographic parameters of $\Lambda CDM$ model.\\
{\bf Critical point $C$}, corresponds to the {\em fluid-kinetic-scaling} solution.
  This solution depends on the coupling constant $\beta$ and exists for all potentials. It has two eigenvalues  depending on both $\alpha$ and $\beta$:
$\lambda_+ = -3/2+\beta^2$ and $\lambda_- = 3+2 \beta^2-2 \beta \alpha$.
The solution is stable  for\\
$ \left\{
   \begin{array}{ll}
   \alpha<\frac{3+2\beta^2}{2\beta}, -\sqrt{\frac{3}{2}}<\beta<0\\  \alpha>\frac{3+2\beta^2}{2\beta},0 <\beta<\sqrt{\frac{3}{2}}  \hbox{$\ \ \ \ \ \ \ \  \ \ \ \ \ \ \ \ $}
   \end{array}
 \right.
$\\
in appropriate phase space shown in Fig.(4). The set of the values of cosmographic parameters at this point is:\\
$ \left\{
   \begin{array}{ll}
   q_{c}=\frac{1}{2}+\beta^{2}\\
   Q_{c}=1+3\beta^{2}+2\beta^{4}\\
   X_{c}=-\frac{7}{2}-\frac{27}{2}\beta^{2}-16\beta^{4}-6\beta^{6}\\
   Y_{c}=\frac{35}{2}+\frac{163}{2}\beta^{2}+134\beta^{4}+94\beta^{6}+24\beta^{8}\\
    \end{array}
 \right.
$\\
 \begin{figure}[t]
\includegraphics[scale=.45]{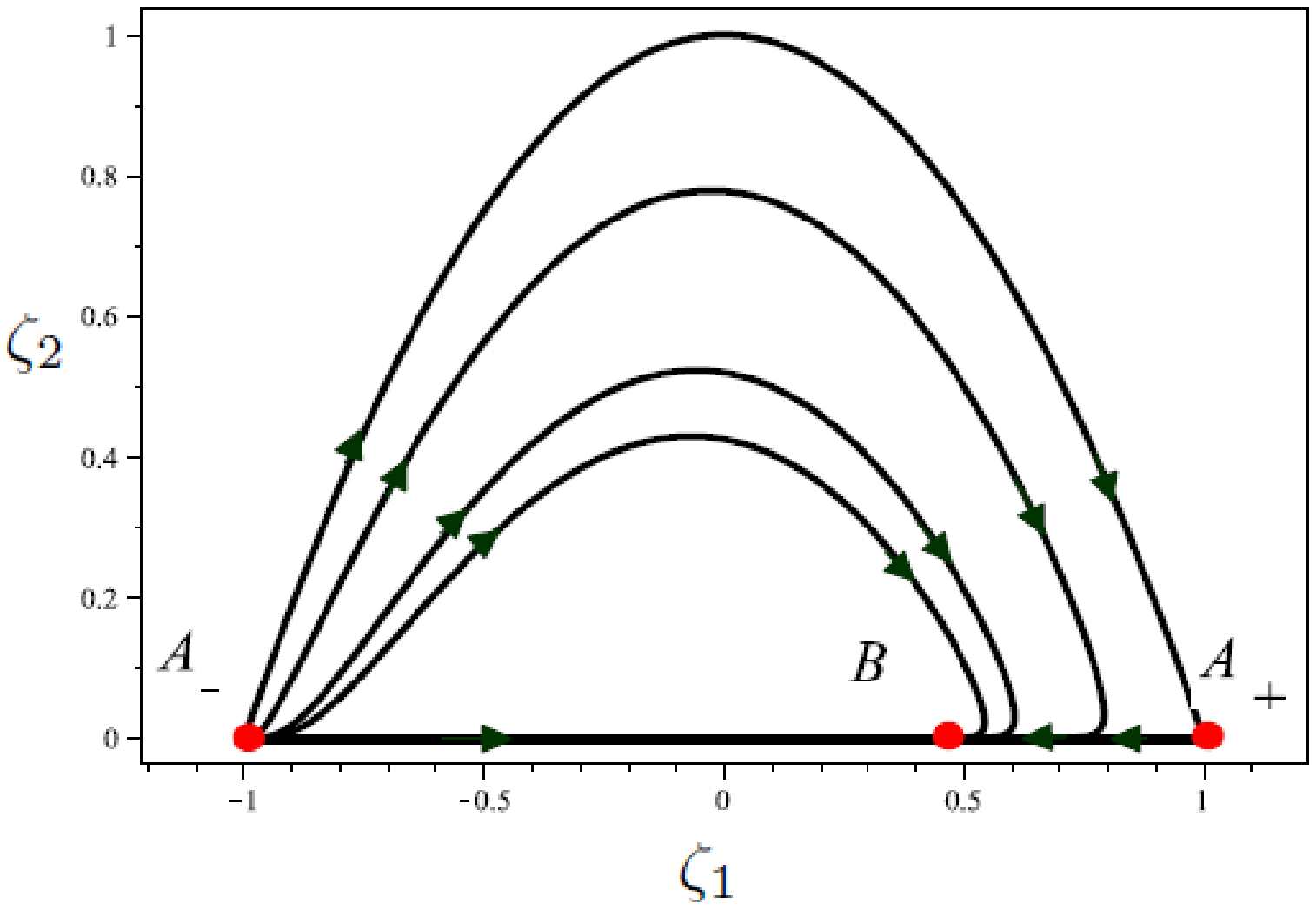}\hspace{0.1 cm}\\
Fig. 3:  The phase space for $\alpha=-1$ and $\beta=-1$ . The late time attractor is scaling solution with $\zeta_{1}=\frac{\sqrt{6}}{6},\zeta_{2}=0$. \\
\end{figure}
 along with chameleon mass $m_{ch}$
\begin{eqnarray}\label{mp1}
\frac{m_{ch}^{2}}{3H^{2}}=\beta^{2}(1-\frac{2\beta^{2}}{3})
\end{eqnarray}
 determining the stability condition $(\beta^{2}<\frac{3}{2})$ for upper bound of chameleon mass in the form of  $m_{ch}^{2}<\frac{9H^{2}}{16}$.
 As $\beta\rightarrow 0$ , the potential tends to $V_{eff}(\phi)\rightarrow \rho_{m}$ and the cosmographic parameters would  be\\ $ q_{c}=1/2,Q_{c}=1,X_{c}=-\frac{7}{2},Y_{c}=\frac{35}{2}$\\
 \begin{figure}[t]
\includegraphics[scale=.45]{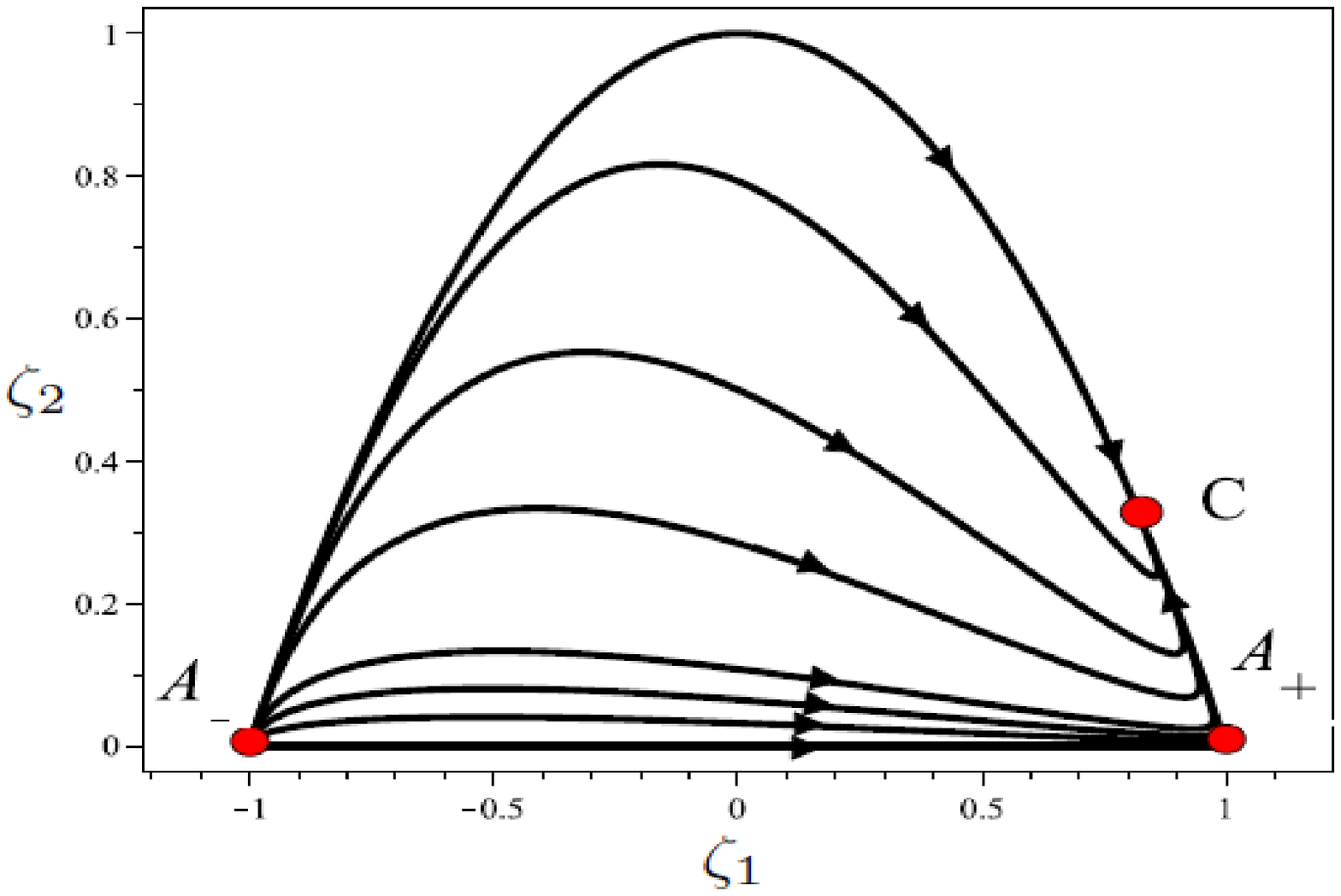}\hspace{0.1 cm}\\
Fig. 4:  The phase space for $\alpha=-4$ and $\beta=-1$ . The late time attractor is scaling solution with $\zeta_{1}=\frac{\sqrt{6}}{3},\zeta_{2}=\frac{1}{3}$.\\
\end{figure}
 which are those corresponding to the matter-dominated era. The radiation-dominated epoch can also be   represented by $\beta^{2}=\frac{1}{2}$,  with $m_{ch}=H$ and cosmographic parameters\\ $q_{c}=1,Q_{c}=3,X_{c}=-15,Y_{c}=105$.\\
  It is worth noting that, while the critical points $B$ and $C$ have the same $Q$ parameter ($ Q_{c}=1$) when $(\alpha,\beta) \rightarrow 0$, however they have different deceleration parameters \\ $point, B:q_{c}=1/2$ and $point, C:q_{c}=-1$.\\
  This points are corresponds to $SCDM$ and $LCDM$ states respectively\cite{Alam}.
  This degeneracy is associated with the order of the derivatives of scale factor. In fact, it is possible to derive $Q_{c}$ in terms of $q_{c}$ as
  \begin{align}
  Q_{c}=2q_{c}^{2}+q_{c}
  \end{align}
  indicating  two possible values of $q_{c}$  for $ Q_{c}>\frac{-1}{8}$ \\
   \begin{align}
  q_{c}=\frac{-1\pm\sqrt{1+8Q_{c}}}{4}
  \end{align}
  \begin{figure}[t]
\includegraphics[scale=.45]{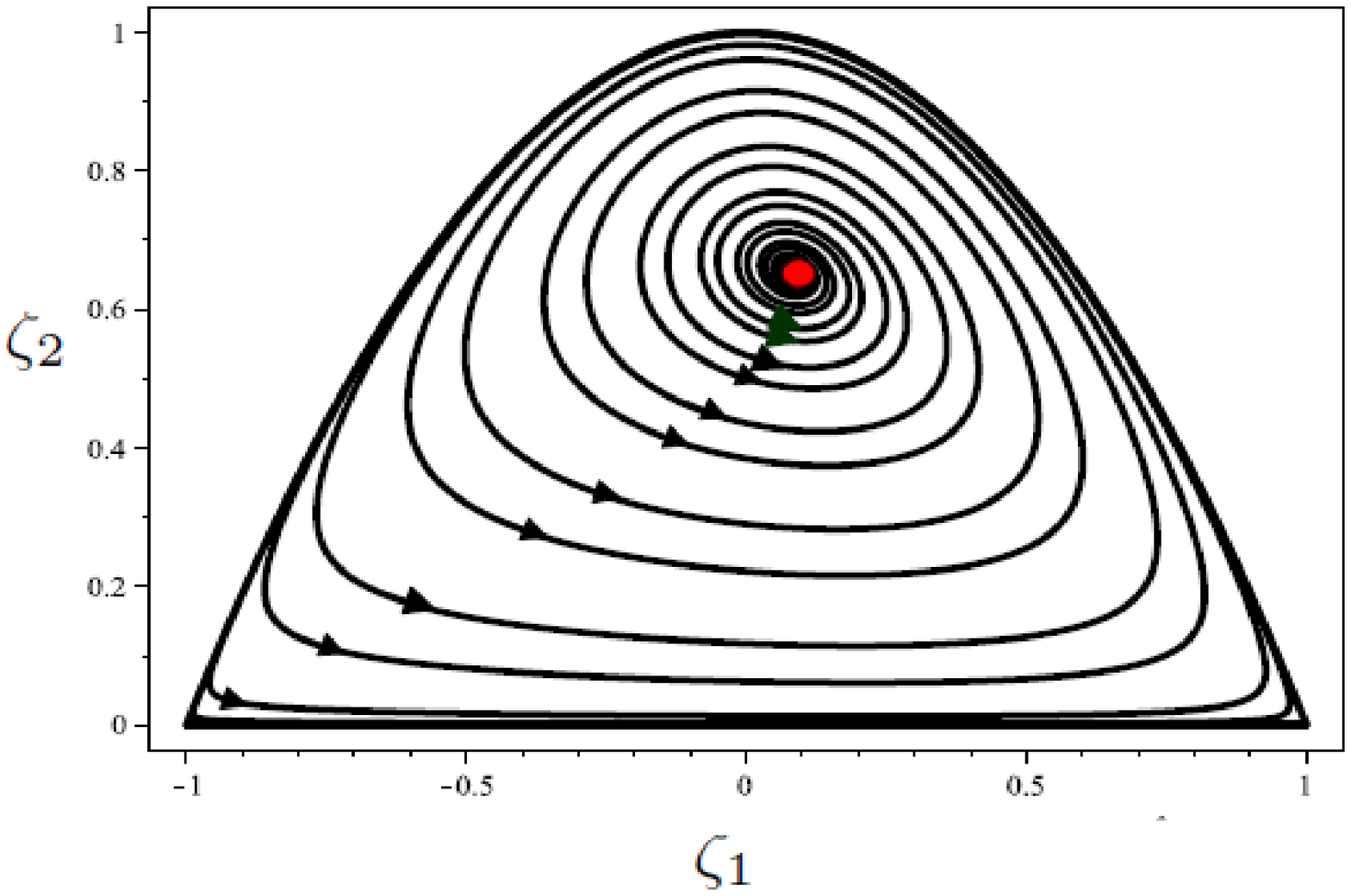}\hspace{0.1 cm}\\
Fig. 5:  The phase space for $\alpha=-10$ and $\beta=4$ . The late time attractor is scaling solution with $\zeta_{1}=\frac{\sqrt{6}}{28},\zeta_{2}=\frac{137}{196}$.\\
\end{figure}
   Meaning that, two different epochs of the universe with different values of the deceleration parameter may have the same parameter $Q$.\\
 {\bf Critical point, $D$} corresponds to a {\em fluid potential-kinetic-scaling} solution with eigenvalues

$ \lambda_+=\frac{\frac{3}{2}\beta-\frac{3}{4}\alpha+\sqrt{A}}{-\beta+\alpha}  \, \qquad \lambda_{-} =\frac{\frac{3}{2}\beta-\frac{3}{4}\alpha-\sqrt{A}}{-\beta+\alpha}  \,\nonumber
$\\
where $A=180\beta^2-108\beta\alpha-63\alpha^2-96\beta^2\alpha^2+48\beta^3\alpha+48\beta\alpha^3+216$.
 The cosmographic parameters at this point are as follows:\\
$ \left\{
   \begin{array}{ll}
   q_{c}=\frac{1}{2}\Bigg(\frac{\alpha+2\beta}{(\alpha-\beta)}\Bigg)\\
   Q_{c}=\frac{1}{2}\Bigg(\frac{5\alpha\beta+2\beta^{2}+2\alpha^{2}}{(\alpha-\beta)^{2}}\Bigg)\\
   X_{c}=-\frac{1}{4}\Bigg(\frac{14\alpha^{3}+39\alpha^{2}\beta+24\alpha\beta^{2}+4\beta^{3}}{(\alpha-\beta)^{3}}\Bigg)\\
   Y_{c}=\frac{1}{4}\Bigg(\frac{70\alpha^{4}+209\alpha^{3}\beta+159\alpha^{2}\beta^{2}+44\alpha\beta^{3}+4\beta^{4}}{(\alpha-\beta)^{4}}\Bigg)\\
      \end{array}
 \right.
$\\
 with chameleon mass $m_{ch}$ given by
\begin{align}\label{mp}
&\frac{m_{ch}^{2}}{3H^{2}}=\alpha^2\Bigg(1-\frac{3}{2(\beta-\alpha)^2}-\frac{(-\beta\alpha+\alpha^2-3)}{(\beta-\alpha)^2}\Bigg)\\
&+\beta^{2}\Bigg(\frac{-\beta\alpha+\alpha^2-3}{(\beta-\alpha)^2}\Bigg)\nonumber
\end{align}
The phase space of system has been shown in Fig.(5), in which the critical point is stable and states follow from initial conditions. It is interesting to note that, the cosmographic parameters at points $C$ and $D$ get the same values when $\beta\rightarrow 0$. Under this condition at these respected points, however, the chameleon mass gets different values as $0$ and $ \frac{9H^{2}}{2}$, respectively. Similarly, for $\alpha\rightarrow 0$, the critical points $B$ and $D$  have the same amounts of cosmographic parameters with different chameleon masses. From equation (\ref{mm2}), because, the chameleon mass $\frac{m_{ch}^{2}}{3H^{2}}$ depends not only on the $(q,\alpha,\beta)$ but also on $x$ or $\zeta_{1}=\frac{x}{\sqrt{6}}=\frac{\phi\dot{}}{\sqrt{6}H M_{pl}}$.

There are the following three solutions for variable $x$, according to the equation (\ref{co}) with $\{q=1/2,Q=1\}$,
\\
 $\left\{
\begin{array}{ll}
x_{1}=0\\
x_{2}=\frac{-3+(9-12\alpha\beta +12\beta^2)^{\frac{1}{2}}}{2(\alpha-\beta)}\\
x_{3}=\frac{-3-(9-12\alpha\beta +12\beta^2)^{\frac{1}{2}}}{2(\alpha-\beta)}\\
\end{array}
\right.
$\\
Where $x_{1}=x_{2}=0$ and $x_{3}=-\frac{3}{\alpha}$  are the critical points $C$ and $D$, by $\beta=0$ constraint. As a result, it is possible to having two or three different critical points with different chameleon mass, while  their cosmographic parameters are the same.\\

\section{Summary and remarks}
In our previous works, the stability analysis of chameleon field and its interaction with other fields have been studied \cite{sal1},\cite{sal3},\cite{sal4}. The chameleon gravity on cosmological scales and constraining its  parameters $(\alpha,\beta)$ have also been investigated for various potentials \cite{sal5}.  This paper has shown that, here, it is possible to reconstruct the chameleon gravity in the presence of exponential potential  $V=M^{4}\exp(\frac{\alpha\phi}{M_{pl}})$  on a cosmological scale, by measuring the first three derivatives of the scale factor $(H,q,Q)$, if the chameleon parameter $\beta$ and coupling constant $\alpha$ are known. In this right, measuring the third derivative of the scale factor $Q$ is of great importance to reconstruct the chameleon free parameters/variables. However, If, nothing could be proposed about the parameters $(\alpha,\beta)$, chameleon model could only be reconstructed by measuring the first five derivatives of the scale factor; as the parameters $(\alpha,\beta)$ will also be determined automatically. If. in addition, we want to test the validity of the chameleon model, the first six derivatives of the scale factor are required. In this respect, we reconstructed the free parameters of the  model based on the best-fitted values of the cosmographic parameters in the significant studies done by the authors of \cite{Tegmark} and \cite{Cap}. In model-reconstructing paradigm using the data referenced to the former, the result was $x_{0}=0$, $y_{0}=0.824,z_{0}=2.176$ and $(\alpha,\beta)\simeq( \pm2.780, \pm 10.01)$; as a case corresponding  to $\frac{\dot{\phi}}{m_{p}H}|_{0}=0$, $\frac{\rho_{m}e^{\frac{\beta\phi}{M_{pl}}}}{3H^{2}M^{2}_{pl}}|_{0}\simeq0.274,z=\frac{V}{3H^{2}M^{2}_{pl}}|_{0}\simeq0.725$. Based on the best-fitted cosmographic values of the latter, the result was $x_{0}=0$, $y_{0}=0.74,z_{0}=.2.262$ and $(\alpha,\beta)\simeq( \pm 2.598, \pm 11.25)$; as a case corresponding to $\frac{\dot{\phi}}{m_{p}H}|_{0}=0$, $\frac{\rho_{m}e^{\frac{\beta\phi}{M_{pl}}}}{3H^{2}M^{2}_{pl}}|_{0}\simeq0.246,z=\frac{V}{3H^{2}M^{2}_{pl}}|_{0}\simeq0.754$.\\
 Recently, some studies have constrained the chameleon parameter $\beta$ by new technics, for example \cite{mart} by searching for (solar) chameleons with the CERN Axion Solar Telescope (CAST), reported as $(1<\beta<10^{6})$. Using neutron interferometry, also, the author of \cite{Lemmel} has found that  the  $\beta$ constant is less than $1.9\times10^{7}$.
   Now, based on the best fisted values of the cosmographic parameters by those mentioned above (\cite{Tegmark},\cite{Cap}), we have found $\beta\simeq10$ and $\beta\simeq11$. These two values are close to each other, lie in the region of those expected by \cite{mart},\cite{Lemmel} and comparable with
   \cite{sal5} and that previously pointed out by Khoury and Weltman in the original chameleon paper \cite{Khourym} (in harmony with  string theory, the $\beta$ parameter should be of the order of unity).
 Finally, we  derived the  value of each of the cosmographic parameters at the level of critical points for chameleon mechanism. The important point was that, all cosmographic parameters can be derived in terms of the deceleration parameter $q$ at these points; meaning  that only deceleration parameter is significant at this level. \\

We have also derived an expression for the chameleon mass in terms of the cosmographic  and chameleon $(\alpha,\beta)$ parameters. The representation of the chameleon mass in terms of the cosmographic parameters shows that, not only the respective field  takes different masses depending on the local matter density but also evolves dynamically in the Universe. It acquires a time-dependent mass by varying at different epochs of the universe. Hence, the field interaction range $\lambda=m_{ch}^{-1}$ is also varying through the space and time.


\begin{thebibliography}{}
\bibitem{Sandage} A. Sandage, Physics Today, February 23, 34 (1970)
\bibitem{Winberg} S. Weinberg, Astrophys. J. Lett., 161, L233 (1970)
\bibitem{Harrison} E.R. Harrison, Nature 260, 591 (1976)
\bibitem{Hut} P.Hut, Nature 267, 128 (1977)
\bibitem[(Reiss 1998)]{Reiss} A.G. Reiss et al, Astron. J. 116, 1009 (1998)

\bibitem[(Bennet 2003)]{Bennet} C. I. Bennet et al, Astrophys J. Suppl. 148:1, (2003)

\bibitem[(Reiss1 1998)]{Reiss1} A. G. Riess, et al, [Supernova Search TeamCollaboration]
Astron J. 116 1009 (1998)
\bibitem[(Pope 2004)]{Pope} A. C. Pope, et. al, Astrophys J. 607 655, (2004)

\bibitem[(Spergel et.al 2003)]{Spergel} D. N. Spergel, et. al., Astrophys J. Supp. 148 175, (2003)

\bibitem[(Sahni 2000)]{Sahni} V. Sahni, A. Starobinsky,  Int. J. Mod. Phys. D 9 373-444, (2000)

\bibitem[(Weinberg 1989)]{Weinberg} S. Weinberg, Rev. Mod. Phys. 61 1(1989)

\bibitem[(Caldwell  \& Dave 1998)]{Caldwell} R. R. Caldwell, R. Dave and P. J. Steinhardt,  Phys. Rev.
Lett. 80 1582,(1998)
\bibitem[(Zlatev et al. 1999)]{Zlatev} I. Zlatev, L. Wang and P. J. Steinhardt, Phys. Rev. Lett. 82
896,( 1999)

\bibitem[(Caldwell et al. 2003)]{Caldwell2} R. R. Caldwell, M. Kamionkowski, N. N. Weinberg, Phys.
Rev. Lett. 91 071301,( 2003)

\bibitem[(Zhu et al. 2004)]{Zhu} Z. H. Zhu, M. K. Fujimoto and X. T. He, Astrophys J.
603 365-370,( 2004)

\bibitem[(Zhu \& Alcaniz 2005)]{Zhu1} Z. H. Zhu and J. S. Alcaniz, Astrophys J. 620 7-11 ( 2005 )


\bibitem[(e.g. Sadeghi et al. 2008, Sadeghi et al. 2009, Guo et al. 2005, Xia et al. 2005, Setare 2006, Zhao \&  Zhang 2006, Zhao et al. 2007, Setare et al. 2008, Setare \& Saridakis 2008,2009)]{Sad08} Sadeghi, J., Setare, M. R., Banijamali, A., \& Milani, F. 2008,  Phys. Lett. B 662 92;
Sadeghi, J., Setare, M. R., Banijamali, A., \& Milani, F. 2009, Phy. Rev. D 79 123003;
Guo, Z. K. et al. 2005, Phys. Lett. B 608, 177; Xia, J.-Q., Feng, B., \&  Zhang, X. 2005,
Mod. Phys. Lett. A 20 2409; Setare, M. R. 2006, Phys. Lett. B 641 130;
Zhao, W. \&  Zhang, Y. 2006, Phy. Rev. D 73  123509; Zhao, G.-B.,  Xia, J.-Q., Feng, B.,
\&  Zhang, X. 2007, Int. J. Mod. Phys. D 16, 1229; Setare,  M. R., Sadeghi, J. \&
Amani, A.R. 2008, Phys. Lett. B 660 299; Setare, M. R. \& Saridakis, E. N. 2008, Phys. Lett.
B 668 177; Setare,  M. R. \&  Saridakis, E. N. 2009, Int. J. Mod. Phys. D 18
549; Setare, M. R. \&  Saridakis, E. N. 2008, J. Cos. Astro. Phys. 09 026.


\bibitem[( Cai et al. 2007)]{Cai07} Cai,  Y. F., Qiu, T., Piao, Y. S., Li,  M. \&  Zhang, X. 2007, JHEP 0710 071.

\bibitem[(Farajollahi, Salehi, Tayebi \& Ravanpak 2011)]{far-salehi}Farajollahi, H., Salehi, A., Tayebi, F., Ravanpak, A. 2011, J. Cos. Astro. Phys. 05, 017.

\bibitem[(Capozziello et al. 2006,2003)]{Cap06} Capozziello, S., Cardone, V. F., Carloni, S. \& Troisi, A. 2006, Int.J.Mod.Phys. D15 69;
2003, Int.J.Mod.Phys. D12 1969.






\bibitem[(Setare 2007)]{Setare} M. R. Setare, Phys. Lett. B644:99-103,(2007)


\bibitem[(Setare \&  Jamil 2010, Davis et al. 2009, Ito \&  Nojiri 2009, Tamaki \& Tsujikawa 2008,Farajollahi \& Salehi 2010b, Mota \& Shaw 2007, Dimopoulos \& Axenides 2005)]{set10} Setare,M. R. \&  Jamil,M. 2010, Phys. Lett. B 690  1-4 ;  Davis,A. C., Schelpe, C. A.O., Shaw, D. J., 2009, Phy. Rev. D 80 064016 ;
 Ito, Y. \&  Nojiri, S. 2009, Phy. Rev. D 79:103008; Tamaki,T. \& Tsujikawa,S. 2008, Phy. Rev. D 78 084028 ; Farajollahi, H. \& Salehi, A. 2010b Int. J. Mod. Phys. D19:621-633;  Mota,D.F. \&  Shaw, D.J. 2007, Phy. Rev. D 75, 063501; Dimopoulos, K. \& Axenides, M. 2005, J. Cos. Astro. Phys. 0506:008.




\bibitem[(Lyth \& Riotto 1999)]{Lyth} D. H. Lyth and A. Riotto, Phys. Rept. 314, 1 (1999)

\bibitem[(Wetterich 1988)]{Wetterich} C. Wetterich, Nucl. Phys. B302, 668 (1988)

\bibitem[(Peebles \& Ratra 1988)]{Peebles} P. J. E. Peebles and B. Ratra, Ap. J. 325 ,L17 (1988)

\bibitem[(Caldwell et al. 1998)]{Caldwell1} R. R. Caldwell, R. Dave and P. J. Steinhardt, Phys. Rev. Lett. 80, 1582 (1998)

\bibitem[(Easson 2007)]{Easson} D. A. Easson, JCAP 070 2, 004 (2007)

\bibitem[(Damouri et. al 1990)]{Damouri} T. Damour, G. W. Gibbons and C. Gundlach, Phys. Rev. Lett, 64, 123 (1990)

\bibitem[(Setare1 et al. 2009)]{Setare1} M. R. Setare, Elias C. Vagenas, Int. J. Mod. Phys. D18:147-157 (2009)

\bibitem[(Carroll 1998)]{Carr} S. M. Carroll, Phys. Rev. Lett. 81 3067(1998)

\bibitem[(carroll et al. 1992)]{carrolll} S. M. Carroll, W. H. Press and E. L. Turner, Ann. Rev. Astron. Astrophys, 30, 499 (1992)

\bibitem[(Biswas et al. 2006)]{Biswass} T. Biswas, R. Brandenberger, A. Mazumdar and T. Multamaki. Phys.Rev. D74 063501, (2006)

\bibitem[(Uzan 2003)]{Uzan} J. P. Uzan, Rev. Mod. Phys. 75, 403 (2003)

\bibitem[(Bertotti et al 2003)]{Bertotti} B. Bertotti et al. Nature 425, 374 (2003)

\bibitem[(Chew \& Frautschi 1998)]{Chew} G. F. Chew and S. C. Frautschi. Phys. Rev. Lett. 7,  394 (1961)

\bibitem[(Damour et al. 2002)]{Damourm} T. Damour, F. Piazza and G. Veneziano, Phys. Rev. D 66 , 046007 (2002)

\bibitem[(Nojiri \& Odintsov 2004)]{nojiri} S. Nojiri, S. D. Odintsov, Mod. Phys. Lett. A 19:1273-1280 (2004)

\bibitem[(Khoury \& Weltman  2004)]{Khoury}J. Khoury and A. Weltman: Phys.Rev.D69:044026,2004;
\bibitem[ (Mota \& Barrow 2004)]{Mota}D. F. Mota, J. D. Barrow, Phys. Lett. B581 141-146(2004);

\bibitem[(Khoury \& Weltman 2004)]{Khourym} J. Khoury and A. Weltman, Phys. Rev. Lett. 93,171104 (2004)

\bibitem[(Brax et al. 2004)]{Brax2} Ph. Brax, C. van de Bruck, A. C. Davis, J. Khoury and A. Weltman. Phys. Rev.D70, 123518 (2004)

\bibitem[(Wetterich 1995)]{Wett} C. Wetterich, Astron. Astrophys. 301, 321 (1995)

\bibitem[(Damour \& Polyakov 1994)]{Dam} T. Damour and A.M. Polyakov, Nucl. Phys. B423, 532 (1994); Gen. Rel. Grav. 26, 1171 (1994)

\bibitem[(Huey et al. 2000)]{Huey} G. Huey, P.J. Steinhardt, B. A. Ovrut and D. Waldram. Phys. Lett. B 476, 379 (2000)

\bibitem[(Hill \&  Ross 1998)]{Hill} C.T. Hill and G. C. Ross, Nucl. Phys. B311, 253 (1988)

\bibitem[(Ellis et al. 1989)]{Ellis} J. Ellis, S. Kalara, K.A. Olive and C. Wetterich, Phys. Lett. B 228, 264 (1989)

\bibitem[(Reiss \& Bruck 2004)]{Mot} D. F. Mota and C. van de Bruck, Astron. Astrophys. 421,71 (2004)

\bibitem{Liske} Liske J. \emph{et al.}, Mon. Not. Roy. Astron. Soc. {\bf386}, 1192 (2008).

 \bibitem {Weinberg2 } S. Weinberg, Gravitation and cosmology: Principles and applications of the general theory of
relativity, (Wiley, New York, 1972).
\bibitem {Visse} M. Visser, Class. Quantum Gravity 21, 2603 (2004)
 \bibitem {Matt Visser} Matt Visser, Gen.Rel.Grav.37:1541-1548,2005
\bibitem {Capozziello} S. Capozziello, V.F. Cardone, V. Salzano,Phys.Rev.D78:063504,2008
\bibitem {Aviles} A. Aviles, C. Gruber, O. Luongo, H. Quevedo, Arxiv:
1204.2007, (2011); A. Aviles, L. Bonanno, O. Luongo, H.
Quevedo, Phys. Rev. D, 84, 103520, (2011).\bibitem{cct} Capozziello, S., Cardone, V.F., Troisi, A. 2005, Phys. Rev. D, 71, 043503
\bibitem{prl} Capozziello, S., Cardone, V.F., Troisi, A. 2006, JCAP, 0608, 001
\bibitem{cap2} S. Capozziello, V. F. Cardone, H. Farajollahi, and A. Ravanpak
Phys. Rev. D 84, 043527
\bibitem {Alam} U. Alam, V. Sahni, T. D. Saini, A. A. Starobinsky, Mon.
Not. Roy. Astron. Soc. 344, 1057, (2003); V. Sahni, T. D.
Saini, A. A. Starobinsky, U. Alam, JETP Lett., 77, 201,
\bibitem{Salehi} A. Salehi, M. R. Setare, A. Alaii, Eur. Phys.J. C 78 (6), 495
\bibitem{Bamba} K. Bamba et al., Astrophys. Space Sci. 342, 155 (2012) [arXiv:1205.3421].
\bibitem{catton} C. Cattoen and M. Visser, Phys. Rev. D 78, 063501 (2008) [arXiv:0809.0537].
\bibitem{Vi} V. Vitagliano, J. Q. Xia, S. Liberati and M. Viel, JCAP 1003, 005 (2010) [arXiv:0911.1249].
\bibitem{catt} C. Cattoen and M. Visser, gr-qc/0703122;
C. Cattoen and M. Visser, Class. Quant. Grav. 24, 5985 (2007) [arXiv:0710.1887];
M. Visser and C. Cattoen, arXiv:0906.5407 [gr-qc].
\bibitem{xu} L. X. Xu and Y. Wang, Phys. Lett. B 702, 114 (2011) [arXiv:1009.0963].
\bibitem{xia} J. Q. Xia et al., Phys. Rev. D 85, 043520 (2012) [arXiv:1103.0378].
[34] M. J. Zhang, H. Li and J. Q. Xia, Eur. Phys. J. C 77, no. 7, 434 (2017) [arXiv:1601.01758].
\bibitem{dun} P. K. S. Dunsby and O. Luongo, Int. J. Geom. Meth. Mod. Phys. 13, 1630002 (2016) [arXiv:1511.06532];
O. Luongo, G. B. Pisani and A. Troisi, Int. J. Mod. Phys. D 26, 1750015 (2016) [arXiv:1512.07076].
\bibitem{avil} A. Aviles, C. Gruber, O. Luongo and H. Quevedo, Phys. Rev. D 86, 123516 (2012) [arXiv:1204.2007];
A. Aviles, A. Bravetti, S. Capozziello and O. Luongo, Phys. Rev. D 87, 044012 (2013) [arXiv:1210.5149];
A. Aviles, A. Bravetti, S. Capozziello and O. Luongo, Phys. Rev. D 87, 064025 (2013) [arXiv:1302.4871];
O. Luongo, Mod. Phys. Lett. A 26, 1459 (2011);
A. de la Cruz-Dombriz et al., JCAP 1612, 042 (2016) [arXiv:1608.03746].
\bibitem{Zhou} Y. N. Zhou, D. Z. Liu, X. B. Zou and H. Wei, Eur. Phys. J. C 76, 281 (2016) [arXiv:1602.07189].
\bibitem{Zou} X. B. Zou, H. K. Deng, Z. Y. Yin and H. Wei, Phys. Lett. B 776, 284 (2018) [arXiv:1707.06367].

\bibitem{Brax1} Ph. Brax et al, JCAP 0411 (2004) 004
\bibitem{cai} Hao Wei, Rong-Gen Cai, Phys.Rev.D71:043504,2005
\bibitem{Brax20} Ph. Brax et al, Phys.Lett.B633:441-446,2006
\bibitem{Baruch} Baruch Feldman, Ann E. Nelson, JHEP 0608 (2006) 002
\bibitem{David} David F. Mota, Douglas J. Shaw,Phys.Rev.Lett.97:151102,2006
\bibitem{Amol} Amol Upadhye, Steven S. Gubser, Justin Khoury, Phys.Rev. D74 (2006) 104024
\bibitem{Brax3} P. Brax, Jerome Martin,Phys.Lett.B647:320-329,2007
\bibitem{Brax5} Ph.Brax, C.van de Bruck, A. -C. Davis, Phys.Rev.Lett.99:121103,2007
\bibitem{Brax6} P. Brax et al, Phys.Rev.D76:085010,2007
\bibitem{Brax8} P. Brax et al, Phys.Rev.D77:015018,2008
\bibitem{Burrage} C. Burrage, Phys.Rev.D77:043009,2008
\bibitem{Salvatore}S. Capozziello, S. Tsujikawa,Phys.Rev.D77:107501,2008
\bibitem{Nelson} A.E.Nelson, J. Walsh, Phys.Rev.D77:095006,2008
\bibitem{Sudipta} S. Das, N. Banerje, Phys.Rev.D78:043512,2008
\bibitem{Chou} A.S.Chou et al. Phys.Rev.Lett.102:030402,2009
\bibitem{Brax7} P. Brax et al. Phys.Rev.D78:104021,2008
 \bibitem{Takashi} T.Tamaki, S.Tsujikawa, Phys.Rev.D78:084028,2008
 \bibitem{Burrage}  C. Burrage, A.C. Davis, Douglas J. ShawPhys.Rev.D79:044028,2009
 \bibitem{Shinji} Shinji Tsujikawa, Takashi Tamaki, Reza Tavakol, JCAP 0905:020,2009
 \bibitem{Nojiri}Yusaku Ito, Shin'ichi Nojiri, Phys.Rev.D79:103008,2009
  \bibitem{Davis} A.C. Davis, Camilla A. O. Schelpe, Douglas J. Shaw Phys.Rev.D80:064016,2009
   \bibitem{bra} P. Brax et al. Phys.Rev.D81:103524,2010
  \bibitem{ca}Camilla A. O. Schelpe, Phys.Rev.D82:044033,2010
 \bibitem{ph} Ph. Brax, C. van de Bruck, A. C. Davis, D. J. Shaw, D. Iannuzzi, Phys.Rev.Lett.104:241101,2010
  \bibitem{brax10}  Philippe Brax, Konstantin Zioutas, Phys.Rev.D82:043007,2010
  \bibitem{ryb} G. Rybka et al. Phys.Rev.Lett.105:051801,2010
  \bibitem{braxx} Ph. Brax, R. Rosenfeld, D. A. Steer, JCAP 1008:033,2010
  \bibitem{set} M. R. Setare, Mubasher Jamil,Phys.Lett.B690:1-4,2010
  \bibitem{brax11} Ph. Brax et al. Phys.Rev.D82:083503,2010
   \bibitem{And} Andrea Zanzi, Phys.Rev.D82:044006,2010
   \bibitem{Levshakov} S. A. Levshakov,Astronomy and Astrophysics, v. 524, A32 (2010)
   \bibitem{Brax13} Philippe Brax, Clare Burrage, Phys.Rev.D82:095014,2010
    \bibitem{Sheykhi} Ahmad Sheykhi, Mubasher Jamil, Phys.Lett.B694:284-288,2011
    \bibitem{faraj} H. Farajollahi, M. Farhoudi, A. Salehi, H. Shojaie, Astrophys.Space Sci.337:415-423, 2012
   \bibitem{Jason} Jason H. Steffen et al. Phys.Rev.Lett.105:261803,2010
   \bibitem{sal1} Hossein Farajollahi, Amin Salehi, JCAP 1011:006,2010
    \bibitem{sal5} H.Farajollahi, Amin salehi, Phys. Rev. D 85, 083514 (2012)
   \bibitem{Radouane} Radouane Gannouji,Phys.Rev.D82:124006,2010
   \bibitem{Radouane} Kurt Hinterbichler, Justin Khoury, Horatiu Nastase, JHEP 1103:061,2011
     \bibitem{Alessandra} Alessandra Silvestri, Phys.Rev.Lett.106:251101,2011
     \bibitem{brax15} Philippe Brax and Guillaume Pignol, Phys. Rev. Lett. 107, 111301
      \bibitem{brax2}   Kurt Hinterbichler, Justin Khoury, Horatiu Nastase, Rogerio Rosenfeld, JHEP 08 (2013) 053
       \bibitem{bisabr} Yousef Bisabr, Phys. Rev. D 86, 127503 (2012)
        \bibitem{Valad} Vladimir Folomeev,  Phys. Rev. D 86, 063008 (2012)
        \bibitem{Amol} Amol Upadhye, Wayne Hu, and Justin Khoury,Phys. Rev. Lett. 109, 041301 (2011)
        \bibitem{John}  John D. Anderson, J. R. Morris, Phys.Rev.D 85, 084017 (2012)
       \bibitem{John2} Johannes Noller,  JCAP07(2012)013
        \bibitem{Ay} Ayumu Terukina, Kazuhiro Yamamoto, Phys. Rev. D 86, 103503 (2012)
         \bibitem{Lucas} Lucas Lombriser, Kazuya Koyama, Gong-Bo Zhao, Baojiu Li, Phys.Rev.D85:124054,2012
         \bibitem{Yu} Yu. N. Pokotilovski,Physics Letters B719 (2013) 341-345
         \bibitem{said} Kh. Saaidi, A. Mohammadi, T. Golanbari, H. Sheikhahmadi, B. Ratra,  Physical Review D 86, 045007(2012)
        \bibitem{said2} Khaled Saaidi, AbolHassan Mohammadi, Physical Review D 85, 023526(2012)
       \bibitem{said3} Khaled Saaidi, Abolhassan Mohammadi, Haidar Sheikhahmadi, Physical Review D 83, 104019 (2011)
      \bibitem{Vladimir} Vladimir Folomeev, Douglas Singleton, Phys. Rev. D 85, 064045 (2012)
      \bibitem{Hees} A. Hees, A. Fuzfa, Phys. Rev. D, 85, 103005, 2012
        \bibitem{Katherine} Katherine Jones-Smith,Phys. Rev. D 85, 043502 (2012)
         \bibitem{bao} Baojiu Li, George Efstathiou, MNRAS, 421, 1431 (2012)
       \bibitem{ph3} Philippe Brax, Axel Lindner, and Konstantin Zioutas, Phys. Rev. D 85, 043014(2012)
        \bibitem{sal3} H. Farajollahi, A. Salehi,JCAP 07(2011)036
       \bibitem{Vladimir2} Vladimir Folomeev, Phys. Rev. D 85, 024008 (2012)
       \bibitem{David} David F. Mota and Camilla A. O. Schelpe, Phys. Rev. D 86, 123002
          \bibitem{Yin} Yin Li, Wayne Hu,Phys. Rev. D 84, 084033 (2011)
    \bibitem{Dzhunushaliev} V. Dzhunushaliev, V. Folomeev, D. Singleton, Phys. Rev. D 84, 084025 (2011)
      \bibitem{Valerio} Valerio Faraoni Phys.Rev.D83:124044,2011
   \bibitem{Katherine} Katherine Jones-Smith, Francesc Ferrer,Phys. Rev. Lett. 108, 221101 (2012)
    \bibitem{sal4} H. Farajollahi, A. Salehi, F. Tayebi, A. Ravanpak,JCAP 05(2011)017
    \bibitem{Jamil} Mubasher Jamil, Ibrar Hussain, D. Momeni,Eur.Phys.J.Plus 126:80,2011
  \bibitem{Brax16} Philippe Brax, Nicola Tamanini, Phys. Rev. D 93, 103502 (2016)
  \bibitem{Ivanov} A.N. Ivanov and M. Wellenzohn,Phys. Rev. D 92, 125004
  \bibitem{Lucila} Lucila Kraiselburd,Phys. Rev. D 97, 104044 (2018)
   \bibitem{Antonio} Antonio Padilla, JCAP03(2016)058
    \bibitem{Ivanov2} A.N. Ivanov and M. Wellenzohn, Phys. Rev. D92, 065006 (2015)
    \bibitem{Lemmel} H. Lemmel,Ph. Brax,A. N. Ivanov,T. Jenke,G. Pignol,Physics Letters B 743 (2015) 310–31
     \bibitem{Clare} Clare Burrage, Jeremy Sakstein, JCAP11(2016)045
     \bibitem{Antonio} Antonio Padilla et al, JCAP03(2016)058
      \bibitem{srael} Israel Quiros, Ricardo Garc?a-Salcedo, Tame Gonzalez, F. Antonio Horta-Rangel,Phys. Rev. D 92, 044055 (2015)
       \bibitem{ata} Attaallah Almasi, Philippe Brax, Davide Iannuzzi, René I. P. Sedmik, Phys. Rev. D 91, 102002 (2015)
        \bibitem{Bri} Bridget Falck, Kazuya Koyama, Gong-bo Zhao, JCAP, 2015, 7, 049
        \bibitem{Anastassopoulos} V.Anastassopoulos,P.L.B,Volume 749,  Pages 172-180 (2015)
          \bibitem{Prolay} Prolay Krishna Chanda, Subinoy Das, Phys. Rev. D 95, 083008 (2017)
       \bibitem{Gergely} L.Gergely, Zoltn Keresztes,Phys. Rev. D 91, 024012 (2015)
         \bibitem{ Baum} S. Baum, G. Cantatore, D. H. H. Hoffmann, M. Karuza, Y. K. Semertzidis, A. Upadhye, K. Zioutas,Physics Letters B 739 (2014) 167
          \bibitem{Surajit} Surajit Chattopadhyay, Antonio Pasqua, Martiros Khurshudyan, Eur. Phys. J. C (2014) 74:3080
            \bibitem{mart} Martin Pernot-Borràs, Joel Bergé, Philippe Brax, Jean-Philippe Uzan, Phys. Rev. D 100, 084006 (2019)
              \bibitem{Lucila} Lucila Kraiselburd, Phys. Rev. D 99, 083516 (2019)
               \bibitem{Taishi Katsuragawa} Taishi Katsuragawa, Tomohiro Nakamura, Taishi Ikeda, Salvatore Capozziello, Phys. Rev. D 99, 124050 (2019)
        \bibitem{Taishi Katsuragawa}  Philippe Brax, Sylvain Fichet,  Phys. Rev. D 99, 104049 (2019)
       \bibitem{Tomohiro Nakamura} Tomohiro Nakamura, Taishi Ikeda, Ryo Saito, Chul-Moon Yoo, Phys. Rev. D 99, 044024 (2019)
    \bibitem{ant} Antonio De Felice, Shinji Mukohyama, Michele Oliosi, Yota Watanabe, Phys. Rev. D 97, 024050 (2018)
     \bibitem{Taishi} Taishi Katsuragawa, Shinya Matsuzaki,Phys. Rev. D 97, 064037 (2018)
     \bibitem{phil} Philippe Brax, Anne-Christine Davis, Rahul Jha, Phys. Rev. D 95, 083514 (2017)
      \bibitem{Haidar} Haidar Sheikhahmadi et al, Eur. Phys. J. C (2019) 79: 1038
     \bibitem{Maciej} Maciej Dunajski, Gary Gibbons,Class. Quant. Grav. 25 (2008)
     \bibitem{Kun} En-Kun Li, Minghui Du, Lixin Xu, MNRAS, 491, 4960 (2020)

\bibitem{Xia} Jun-Qing Xia, Vincenzo Vitagliano, Stefano Liberati, and Matteo Viel, Phys. Rev. D. 85, 043520 (2012)
\bibitem{Tegmark} M. Tegmark et al., Sloan Digital Sky Survey Collaboration. Phys.Rev. D 69, 103501 (2004)
\bibitem{Cap} S. Capozziello, V.F. Cardone, H. Farajollahi, A. Ravanpak, Phys.Rev. D 84, 043527 (2011)

\end{thebibliography}
\end{document}